# Interacting electrodynamics of short coherent conductors in quantum circuits

C. Altimiras, F. Portier, P. Joyez[A]

*SPEC, CEA, CNRS, Université Paris-Saclay*
*CEA-Saclay 91191 Gif-sur-Yvette, France*

(Dated: April 27, 2016)

**Abstract** When combining lumped mesoscopic electronic components to form a circuit, quantum fluctuations of electrical quantities lead to a non-linear electromagnetic interaction between the components that is not generally understood. The Landauer-Büttiker formalism that is frequently used to describe non-interacting coherent mesoscopic components is not directly suited to describe such circuits since it assumes perfect voltage bias, i.e. the absence of fluctuations. Here, we show that for short coherent conductors of arbitrary transmission, the Landauer-Büttiker formalism can be extended to take into account quantum voltage fluctuations similarly to what is done for tunnel junctions. The electrodynamics of the whole circuit is then formally worked out disregarding the non-Gaussianity of fluctuations. This reveals how the aforementioned non-linear interaction operates in short coherent conductors: voltage fluctuations induce a reduction of conductance through the phenomenon of dynamical Coulomb blockade but they also modify their internal density of states leading to an additional electrostatic modification of the transmission. Using this approach we can account quantitatively for conductance measurements performed on Quantum Point Contacts in series with impedances of the order of $R_K = h/e^2$. Our work should enable a better engineering of quantum circuits with targeted properties.

## I. INTRODUCTION : INTERACTIONS IN QUANTUM CIRCUITS

At small scales and low temperatures, electronic components become quantum: their state is not described anymore by classical currents and voltages, but by operators which have quantum fluctuations. When considering several of these components interconnected at a scale larger than the electronic coherence length (so that electronic interferences between components vanish) one recovers the familiar lumped-element description of the whole circuit, just like taught in high school for classical electrical circuits (Fig. 1a). In such a circuit, the Kirchhoff laws apply to the operator-valued currents and voltages, including their quantum fluctuations, at all frequencies for which the lumped description applies. The enforcement of these laws requires that the various branches of the circuit accommodate for the presence of each other. This accommodation can be seen as a fluctuation-mediated electromagnetic interaction that is non-local in space and frequency, rendering the system generally non-linear. In practice, this non-linear interaction is understood quantitatively only in a few, restricted quantum circuits. The aim of the present work is to reach a more general understanding of how this interaction operates. For sake of simplicity we only consider circuits in which neither static charging effects nor Kondo effect occur (semi-isolated "islands" shall have either negligible or extremely large charging energy). Let us start by reviewing the different levels at which this interaction has been taken into account and understood in lumped quantum circuits so far.

### A Different states of consideration and handling of the interactions in lumped quantum circuits

In the case of nearly non-dissipative circuits (such as e.g. qubit circuits), one can generally write a Hamiltonian and solve the Schrödinger equation to work out this complex electromagnetic interaction implicitly and globally, sparing the need to understand how it operates in detail. In contrast, in open quantum circuits biased out-of-equilibrium no such global solving method exists, and they were up to now considered mostly in restrictive situations where such a fully developed interaction between various parts of a circuit does not occur.

[A] email: philippe.joyez@cea.fr

In particular, when investigating an individual quantum component, it is most frequently connected to macroscopic leads which naturally nearly implement an ideal (i.e. classical, fluctuationless) voltage-bias situation in which the Landauer-Büttiker (LB) scattering formalism [1, 2] can be used. However knowing (theoretically or experimentally) the behavior of the component only under this situation is not sufficient to predict its behavior when inserted in an arbitrary quantum circuit.

A case where voltage fluctuations have a well understood effect on a quantum component is that of a tunnel junction. The junction indeed becomes non-linear in presence of voltage fluctuations, with its conductance at low voltages that can be strongly reduced. This phenomenon is known as Dynamical Coulomb Blockade (DCB) and is quantitatively explained by the $P(E)$ theory [3, 4, 5]. In this theory however, the tunnel current through the junction being weak, the junction's back-action on the rest of the circuit is disregarded. The $P(E)$ theory hence considers only an unidirectional interaction (i.e. without the "reaction" part), but the phenomenon of DCB itself can be seen as the generic signature of the interaction. The case of a channel of arbitrary transmission where voltage fluctuations are small (so that conductance changes remain small too) is also well understood [6, 7, 8, 9], with predictions similar to the tunnel junction case but with the reduction of the conductance multiplied by the channel's Fano factor.

Finally situations where full fluctuation-mediated interaction (with action and reaction) takes place are understood only in few cases: (i) for a single channel of arbitrary transmission connected to a pure resistor, which has been addressed by a wide range of techniques with more or less restrictive hypotheses [10, 11, 12, 13, 14] and (ii) in the case of a resonant level in series with resistors [15, 16, 17]. In Ref. [18] it was shown that both cases can be well accounted for by using a mapping of these systems onto the physics of an impurity in a Tomonaga-Luttinger Liquid (TLL). This mapping predicts scaling laws for the conductance with a characteristic energy scale that is however not predicted. Importantly, there are no theories readily applicable for non-ohmic environments such as high impedance resonant structures that were recently used in experiments probing the radiative signatures of DCB [19, 20, 21].

### B  Outline of the article

In this article we work out how the fluctuation-mediated interaction occurs in a somewhat general dissipative circuit consisting of a Short Coherent Component (SCC) that can be described in the LB formalism and an arbitrary external circuit with substantial quantum fluctuations. In Sec. II., we first show that the usual LB formalism that describes ideal voltage-biased situations can be simply extended in the spirit of the $P(E)$ theory, to take into account quantum voltage fluctuations that are slower than the traversal time of electrons through the coherent component. We show and discuss how quantum fluctuations modify qualitatively the current noise and the admittance of the SCC predicted in the LB formalism. In a second step we show that by disregarding the non-Gaussianity of all fluctuations and treating them on an equal footing, the electrodynamic response of the SCC and the fluctuations in the circuit can be formally worked out simultaneously in a self-consistent manner, in presence of arbitrarily strong interaction and including in nonequilibrium situations. Our synthesis of LB and $P(E)$ theories clearly pinpoints how the electromagnetic interaction takes place, acting in particular on the internal charge degree of freedom in the SCC arising from finite dwell time.

In Sec. III. we illustrate the use of our formal general solution, and show that the predictions obtained in this approach are able to account for measurements performed on Quantum Point Contacts in 2DEGs with arbitrary transmission, with minimal assumptions on the experiments.

## II.  FORMAL SOLUTION OF INTERACTING TRANSPORT IN SHORT COHERENT CONDUCTORS

### A  Description of the system

We consider a two-terminal SCC, embedded in a circuit that has substantial voltage fluctuations (Fig. 1a). This circuit is fairly general and may include conventional linear components such as inductors, capacitors and resistors as well as nonlinear quantum components such as, for instance, other SCCs (that would be simultaneously treated by the method we describe here). We assume that the leads that interconnect the various components are longer than the coherence length so that electronic interferences do not occur between the various components, but still short enough to neglect propagation effects at the frequencies we consider. We know that electrical transport in such coherent conductor taken isolatedly is well described using the Landauer-Büttiker (LB) scattering formalism [22] relevant for non-interacting electrons. Our aim is to extend the LB description of quantum components in order to take into account electromagnetic fluctuations in their surrounding circuit and their interaction with electrical transport. To this end, each node in the circuit is considered as an LB electronic reservoir whose potential has quantum fluctuations and which feeds the components with its local electronic distribution function. For simplicity we assume the short conductor has only two leads, but our approach is easily generalized to several leads.

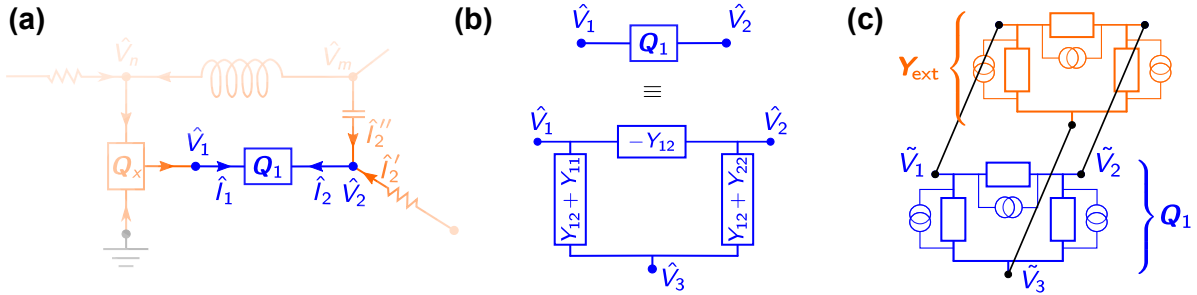

FIG. 1. (a) Schematics of an hypothetical quantum circuit where electrical quantities are operators with quantum fluctuations. We single-out a mesoscopic 2-terminal electronic component $Q_1$ in the circuit. The Landauer-Büttiker formalism is extended to describe transport in the component, taking into account voltage fluctuations. This electron scattering approach predicts that the component generates current fluctuations which the whole circuit transforms into matching voltage fluctuations. (b) The energy dependence of the scattering matrix implies the component has an implicit internal node with an associated degree of freedom ($\hat{V}_3$) whose dynamics needs to be taken into account. The small ac-signal electrodynamics of the component is described by an admittance matrix which depends not only on the scattering matrix of $Q_1$ but also on the voltage fluctuations across $Q_1$ that are determined by the whole circuit. The drawing shows a possible lumped element equivalent circuit of the admittance matrix [23]. (c) Disregarding non-Gaussianity, current and voltage fluctuations at a given bias point are related simply by the small ac-signal (locally linear) electrodynamics of the full circuit. As seen from $Q_1$, the rest of the circuit is also described by an admittance matrix $Y_{\text{ext}}$ that includes in particular all geometrical capacitances. The voltage fluctuations $\tilde{V}_1$, $\tilde{V}_2$, $\tilde{V}_3$ across $Q_1$ result from the noise sources in the whole circuit (drawn here as current sources), all treated on the same footing. Depending on the circuit details, the lumped elements depicted here may depend on the dc bias condition, resulting in a globally non-linear response of the system.

### B  Standard description of transport in the LB formalism

Our starting point is the standard expression of the current in lead $m=1,2$ of the SCC in the Landauer-Büttiker scattering formalism as formulated in Ref. [24]

$$I_m(t) = \frac{e}{h} \int dE dE' [\boldsymbol{a}_m^\dagger(E)\boldsymbol{a}_m(E') - \boldsymbol{b}_m^\dagger(E)\boldsymbol{b}_m(E')] e^{i(E-E')t/\hbar}$$

where $\boldsymbol{a}_m = (a_{m1} \ldots a_{mM})$ and $\boldsymbol{b}_m = (b_{m1} \ldots b_{mM})$ are vectors of operators which respectively annihilate incoming carriers and outgoing carriers in lead $m$, and where $j = 1 \ldots M$ indexes the channels we consider in the leads (each spin direction counting for its own channel). What these operators describe are precisely Landau's Fermi liquid quasiparticles. They obey the relations:

$$\begin{aligned}
\{a_{ij}(E), a_{kl}^\dagger(E')\} = \{b_{ij}(E), b_{kl}^\dagger(E')\} &= \delta_{ij}\delta_{kl}\delta(E-E') \\
\{a_{ij}(E), a_{kl}(E')\} = \{b_{ij}(E), b_{kl}(E')\} &= 0 \\
\langle \boldsymbol{a}_i^\dagger(E)\boldsymbol{a}_j(E')\rangle = \text{Tr}(\rho \boldsymbol{a}_i^\dagger(E)\boldsymbol{a}_j(E')) &= \delta_{ij}\delta(E-E')f_i(E)I_M
\end{aligned}$$

where $\delta$ is here the Kronecker symbol, $I_M$ is the identity matrix of size $M$, $\rho$ the density matrix and $f_m(E)$ is the quasiparticle distribution function in the reservoirs $m=1,2$ which is not necessarily thermal. The outgoing operators are connected to the incoming operators by the usual scattering matrix $\mathcal{S}$ with dimensions $2M \times 2M$

$$\begin{bmatrix} \boldsymbol{b}_1 \\ \boldsymbol{b}_2 \end{bmatrix} = \mathcal{S} \cdot \begin{bmatrix} \boldsymbol{a}_1 \\ \boldsymbol{a}_2 \end{bmatrix}$$

defined in absence of the external circuit. Then, we can express the currents using only incident modes

$$I_m(t) = \frac{e}{h} \int dE dE' [\mathbf{a}_1^\dagger(E) \mathbf{a}_2^\dagger(E)] \cdot A_m(E, E') \cdot \begin{bmatrix} \mathbf{a}_1(E') \\ \mathbf{a}_2(E') \end{bmatrix} e^{i(E-E')t/\hbar} \qquad (1)$$

with the matrix $A_m$ expressed in terms of the $M \times M$ submatrices of $\mathcal{S}$

$$A_m(E, E') = \begin{pmatrix} I_M \delta_{m1} - \mathcal{S}_{m1}^\dagger(E) \mathcal{S}_{m1}(E') & -\mathcal{S}_{m1}^\dagger(E) \mathcal{S}_{m2}(E') \\ -\mathcal{S}_{m2}^\dagger(E) \mathcal{S}_{m1}(E') & I_M \delta_{m2} - \mathcal{S}_{m2}^\dagger(E) \mathcal{S}_{m2}(E') \end{pmatrix}. \qquad (2)$$

The average current in lead 1 is then

$$\langle I_1(t) \rangle = \langle I_1 \rangle = \frac{e}{h} \int dE \, (I - \mathcal{S}_{11}^\dagger(E) \mathcal{S}_{11}(E)) f_1(E) - \mathcal{S}_{12}^\dagger(E) \mathcal{S}_{12}(E) f_2(E)$$

$$= \frac{e}{h} \int dE \, (f_1(E) - f_2(E)) T(E) \qquad (3)$$

with $T(E) = \text{Tr} \, \mathcal{S}_{12}^\dagger(E) \mathcal{S}_{12}(E) = \sum_{n=1}^{M} T_n(E)$ (each spin direction counts for a channel) the total transmission of the SCC that does not depend on the direction considered. The dc current in lead 2 is obtained by exchanging indices 1 and 2 and is therefore opposite, due to the orientation convention chosen: $\langle I_2 \rangle = -\langle I_1 \rangle$. However, this relation for the average values of the currents does not hold for instantaneous values because the operators $I_1$ and $-I_2$ defined by Eq. (1) are different. This can be traced to an internal (charge) degree of freedom in the SCC, whose density of states is due to energy dependence of the scattering matrix. Indeed, the "differential mode" current $\frac{1}{2}(I_1 - I_2)$ can be viewed as the current actually flowing through the SCC, while the "common mode" current $I_1 + I_2$ corresponds to a charge accumulation on an implicit internal node of the SCC (See Fig. 1b, Fig. 2) that can be seen as one electrode of the "quantum capacitance" of the SCC introduced by Büttiker [25, 26, 27]. This internal degree of freedom can have a significant impact on the SCC dynamics, leading to non-intuitive behavior (see e.g. Ref. [28]) because we are all familiar with current-conserving electronic components. In order to unravel the complexity brought by this internal dynamics it is convenient in the following to explicitly add a third terminal corresponding to the internal node of the SCC and in which flows the (displacement) current $I_3(t) = -I_1(t) - I_2(t)$ (with the associated matrix $A_3 = -A_1 - A_2$). This third terminal is at an electrostatic potential $V_3$ and is connected to the rest of the circuit only through geometrical capacitances that, in our approach, are considered part of the external circuit. In this hereafter three-terminal device, gauge-invariance is respected, charge accumulation does not occur anymore, which enables us to use the usual circuit reasoning and techniques. Provided one adds the relevant geometrical capacitances in the external circuit this will also properly take into account screening effects [26, 29] in the limit where the SCC's internal state can be described by a single potential.

### C  Adding voltage fluctuations to Landauer-Büttiker formalism

*Fluctuations of reservoir potentials*

Now we incorporate the voltage fluctuations across the SCC which are due to the electromagnetic degrees of freedom in the circuit. We separate the potential of each reservoir into a dc potential $V_m$ ($m = 1, 2$) and a fluctuating part $\tilde{V}_m(t)$. The dc potential is incorporated in the electrochemical potential, while the fluctuating part leads to a time-dependent phase $\varphi_m(t) = \frac{e}{\hbar} \int_{-\infty}^{t} \tilde{V}_m(\tau) d\tau$ ($m = 1, 2$), which is shared by all electrons in a given reservoir. Even though the quasiparticles are non-interacting, they are related to the original electrons and have the same electrochemical potential; they are hence influenced by the potentials just like the electrons. We then rewrite Eq. (1) as

$$I_m(t) = \frac{e}{h} \int dE dE' [\mathbf{a}_1^\dagger(E) e^{-i\varphi_1(t)} \mathbf{a}_2^\dagger(E) e^{-i\varphi_2(t)}] \cdot A_m(E, E') \cdot \begin{bmatrix} \mathbf{a}_1(E') e^{i\varphi_1(t)} \\ \mathbf{a}_2(E') e^{i\varphi_2(t)} \end{bmatrix} e^{i(E-E')t/\hbar}$$

which can be reverted to the original form of Eq. (1), but with a modified $\mathcal{S}$ matrix (see Eq. (4) below) where the anti-diagonal blocks pick a global phase $\pm i\,\varphi(t)$ where $\varphi(t) = \varphi_2(t) - \varphi_1(t)$ is the time integral of the fluctuating voltage $\tilde{V}(\tau) = \tilde{V}_2(\tau) - \tilde{V}_1(\tau)$ across the SCC. In this writing, the effect of fluctuations of the reservoir voltages is simply to add a random phase factor to the scattering matrix elements that correspond to transferring a quasiparticle from one lead to the other, otherwise leaving the modulus of the scattering amplitudes *a priori* unchanged (this point is discussed more in depth below). This phase factor is the same as that appearing in the tunneling Hamiltonian used in the $P(E)$ theory and can be seen as a charge translation operator [5] which tightly couples the transfer of an electrical charge $e$ in the electromagnetic circuit to the quasiparticle transfer in the scatterer. Introducing such a phase factor is also closely related to what is done in presence of a classical ac drive [30, 31, 32] and likewise it is reasonable if the typical dwell time $\tau$ of the quasiparticles in the SCC is fast compared to the period of the ac drive or, here, the relevant timescale of phase fluctuations in the electromagnetic environment.

*Fluctuations of the internal potential*

Much like the potential of the reservoirs, the potential $V_3$ of the internal node may also fluctuate under the influence of the stochastic scattering of electrons in the SCC, or by capacitive coupling to external signals. This effect was previously considered in e.g. Ref. [29] in the case of a QPC, where it was shown that fluctuations of the internal potential (assumed uniform) induces a phaseshift $\phi(t) \equiv \varphi_3(t) = \frac{e}{\hbar} \int_{-\infty}^{t} \tilde{V}_3(\tau) d\tau$ that comes in factor of the whole scattering matrix.

*Modified scattering matrix*

Finally the full modified scattering matrix that minimally takes into account all potential fluctuations writes

$$\tilde{\mathcal{S}}(E,t) = e^{i\phi(t)} \begin{pmatrix} \mathcal{S}_{11}(E) & \mathcal{S}_{12}(E)\,e^{i\varphi(t)} \\ \mathcal{S}_{21}(E)\,e^{-i\varphi(t)} & \mathcal{S}_{22}(E) \end{pmatrix} \tag{4}$$

Here, the presence of both energy and time arguments in $\tilde{\mathcal{S}}$ corresponds to our separation between static quantities ($E$) and dynamic quantities ($t$). This decoupling of timescales is the first key hypothesis we make in this work. Otherwise one would need to consider the dynamics of the electrons crossing the SCC while interacting with a time-varying field [33] and the modification of the scattering matrix would depend both on the internal structure of the scatterer and on the detailed dynamics of phase fluctuations, breaking the simple and universal factorization of phase fluctuations we consider here in $\tilde{\mathcal{S}}$. The relevant dwell time in the SCC can be estimated as the Wigner-Smith time $\tau \sim \max \hbar \frac{dT_n(E)}{dE}$. Note that the simple evaluation of the dc currents as $\langle I_m(t) \rangle$ (Eq. (3)) in the scattering formalism is determined by the transfer probabilities of independent single electrons between the leads for which the phase factors just introduced have no effect. However, after an electron is transferred, the transfer of a subsequent electron is affected by the circuit's transient (phase) response to the current pulse of the first electron, and this modifies the dc current. We now focus on these correlation effects at the heart of the DCB effect.

### D  Current fluctuations and admittance : Dynamics of the SCC

Let us first consider the $3 \times 3$ non-symmetrized stationnary current correlations [2, 34, 22] matrix $\boldsymbol{S}_{II}(t)$ whose elements are given by

$$(\boldsymbol{S}_{II}(t))_{mn} = S_{I_m I_n}(t) = \langle I_m(t) I_n(0) \rangle - \langle I_m \rangle \langle I_n \rangle \qquad m,n \in \{1,3\}$$

At most three elements of $\boldsymbol{S}_{II}(t)$ are independent since they obey $S_{I_m I_n}(t) = (S_{I_n I_m}(-t))^*$ [35] and the correlators involving $I_3$ are trivially related to those involving only $I_1$ or $I_2$. Carrying out the evaluation similarly to Refs. [2, 34, 22] the current correlators can be expressed as

$$S_{I_m I_n}(t) = \frac{e^2}{\hbar^2} \int d\varepsilon d\varepsilon'\, e^{i(\varepsilon - \varepsilon')t} \operatorname{Tr}\left[\left\langle \begin{pmatrix} f_1(\varepsilon) & 0 \\ 0 & f_2(\varepsilon) \end{pmatrix} . \tilde{A}_m(\varepsilon, \varepsilon', t) . \begin{pmatrix} f_{-1}(\varepsilon') & 0 \\ 0 & f_{-2}(\varepsilon') \end{pmatrix} . \tilde{A}_n(\varepsilon', \varepsilon, 0) \right\rangle_{\text{em}} \right] \tag{5}$$

where $\tilde{A}_m(E, E', t)$ is the matrix $A_m(E, E')$ of Eq. (2) with $\tilde{\mathcal{S}}(\ldots, t)$ replacing $\mathcal{S}(\ldots)$, and $\langle\ldots\rangle_{\text{em}}$ denotes averaging over the electromagnetic degrees of freedom. We also introduced the short-hand notation for the complementary fermionic distribution functions in the reservoirs $f_{-i}(\varepsilon) \equiv 1 - f_i(\varepsilon)$. In order to clarify general features of the above matrix expression, let us explicit one element as an example:

$$S_{I_1 I_2}(t) = -\frac{e^2}{\hbar^2} \sum_{n=1}^{M} \int d\varepsilon\, d\omega\, e^{-i\omega t}\ e^{J_+(t)} f_1(\varepsilon) t_n(\varepsilon) r_{1n}^*(\varepsilon) f_{-2}(\varepsilon+\hbar\omega) t_n(\varepsilon+\hbar\omega) r_{2n}^*(\varepsilon+\hbar\omega) +$$
$$e^{J_-(t)} f_2(\varepsilon) t_n^*(\varepsilon) r_{2n}(\varepsilon) f_{-1}(\varepsilon+\hbar\omega) r_{1n}(\varepsilon+\hbar\omega) t_n^*(\varepsilon+\hbar\omega) +$$
$$f_2(\varepsilon) t_n^*(\varepsilon) f_{-2}(\varepsilon+\hbar\omega) t_n(\varepsilon+\hbar\omega)(r_{2n}(\varepsilon) r_{2n}^*(\varepsilon+\hbar\omega) - 1) +$$
$$f_1(\varepsilon) t_n(\varepsilon) f_{-1}(\varepsilon+\hbar\omega) t_n^*(\varepsilon+\hbar\omega)(r_{1n}^*(\varepsilon) r_{1n}(\varepsilon+\hbar\omega) - 1) \quad (6)$$

where $r_{1n} = (\mathcal{S}_{11})_{nn}$, $r_{2n} = (\mathcal{S}_{22})_{nn}$, are the reflection amplitude in the $n$th channel when respectively arriving from lead 1 or 2, and $t_n = (\mathcal{S}_{12})_{nn}$ the transmission amplitude. We also introduced the notations

$$e^{J_\pm(t)} = \langle e^{\pm i\varphi(t)} e^{\mp i\varphi(0)}\rangle_{\text{em}} \quad (7)$$

for the phase correlators. It is a general feature of all the correlators $S_{I_m I_n}(t)$ that the $e^{J_\pm(t)}$ factors appear only in terms having Fermi factors $f_{\pm 1} f_{\mp 2}$ that hence correspond to the transfer of an electron from one reservoir to the other, while the terms $f_{\pm 1} f_{\mp 1}$ and $f_{\pm 2} f_{\mp 2}$ correspond to 2nd order scattering processes in which an electron returns after a short while to its initial reservoir but at a different energy. Note that at this level the fluctuations of the internal potential simply vanish.

From the current correlators, using the linear response theory [36, 35], we formally get the admittance matrix

$$Y(\omega) = \frac{1}{\hbar\omega} \int \theta(t)\, 2\, S_{II}^-(t)(e^{i\omega t} - 1)\frac{dt}{2\pi} \quad (8)$$

that describe the small-AC signal electrodynamics of the SCC around its dc bias point: for a vector of infinitesimal excitations $[dV_1(\omega), dV_2(\omega), dV_3(\omega)]$, the response currents are given by $[dI_1(\omega), dI_2(\omega), dI_3(\omega)] = Y(\omega).[dV_1(\omega), dV_2(\omega), dV_3(\omega)]$ [37]. In Eq. (8), $S_{II}^-(t) = \frac{1}{2}(S_{II}(t) - S_{II}(-t)) = i\,\text{Im}\, S_{II}(t)$ is the time-antisymmetric part of the current correlator and $\theta$ the Heaviside function. The symmetries of the current correlators imply $Y_{ij}(\omega) = Y_{ji}(-\omega)^* = Y_{ji}(\omega)$. Let us stress that although the linear response theory Eq. (8) is usually invoked in equilibrium situations, it remains rigorously valid for nonlinear systems in arbitrary stationary out-of-equilibrium situations [38, 35] (e.g. in presence of a finite dc bias voltage $V = V_1 - V_2$), provided all quantities entering Eq. (5) (and thus Eq. (6)) are taken in the out-of-equilibrium state. The nonlinear $I - V$ characteristics of the SCC is obtained by integrating the differential conductance (e.g. $Y_{11}(\omega=0)$) w.r.t. the bias voltage. The locally linear electrodynamic behavior of the SCC described by this admittance matrix can be represented using the 3-components lumped model [23] shown in Fig. 1b, as a three-terminal charge-conserving component. Let us finally remind that the scattering matrix describes individual quasiparticle transfers but not displacement currents. Hence the electrodynamic model of the SCC we have just derived from the scattering matrix must generally be complemented by additional geometrical capacitances present in the device and which here, for convenience, we incorporate in the description of the external circuit (see below).

### 1. Differences with the standard results of the LB formalism

In absence of voltage fluctuations due to an external circuit, the phase correlators have no effect ($e^{J_\pm(t)} = 1$) and from Eqs. (5) and (8) one recovers the known expressions for the noise and admittance in the LB formalism [22, 39]. In particular, in that case, Büttiker, Prêtre and Thomas gave a particularly simple and compact expression of the equilibrium admittance (Eq. (2) in Ref. [24])

$$Y_{11}(\omega) = G_K \sum_{n=1}^{M} \int d\varepsilon \frac{f(\varepsilon) - f(\varepsilon+\hbar\omega)}{\hbar\omega}(1 - r_{1n}^*(\varepsilon) r_{1n}(\varepsilon+\hbar\omega))$$
$$Y_{12}(\omega) = -G_K \sum_{n=1}^{M} \int d\varepsilon \frac{f(\varepsilon) - f(\varepsilon+\hbar\omega)}{\hbar\omega} t_n^*(\varepsilon) t_n(\varepsilon+\hbar\omega) \quad (9)$$

From this, one can show that at frequencies $\omega \lesssim \tau^{-1}$ the dynamics of the SCC is simple: in addition to the dc conductance of the SCC determined by its total channel transmission (i.e. the "Landauer formula"), the admittance has an additional imaginary term $iE\omega$ proportional to frequency, where $E$ is called the "emittance" of the SCC [27], related to a partial internal DOS.

In presence of voltage fluctuations, however, the presence of the $e^{J_\pm(t)}$ phase correlators in the elements of $S_{II}(t)$ (e.g. Eq. (6)) will imprint the dynamics of the electromagnetic environment onto the SCC, and make our predictions regarding the system differ qualitatively from the above standard LB results. Notably, the zero-bias conductance of the SCC (given by Eq. (8) for $\omega = 0$ and $V_1 - V_2 = 0$) cannot be identified anymore with its total transmission, resulting in a first key message: the "Landauer formula" is *not* generally valid in quantum circuits, because of voltage fluctuations. Also, the above simple Eqs. (9) for the admittance cannot be simply modified in presence of voltage fluctuations because matrix symmetries assumed in their derivation (for details see [2]) are broken by the presence of the phase correlators. In short, just like for a tunnel junction, in presence of voltage fluctuation, transport in the SCC can no longer be regarded as an internal property; it depends non-trivially on the external circuit through the phase correlator, and this is a second key message of our work.

Interestingly, in our approach, one distinguishes two types of contributions to the elements of the noise and admittance matrices: those having factors $e^{J_\pm(t)}$ which are hence affected by voltage fluctuations and the others which are insensitive to fluctuations. Thus, each element of the admittance matrix of a SCC (or each element of the lumped model equivalent shown in Fig. 1b) can be formally viewed as two well identified independent "components" connected in parallel:

- one affected by voltage fluctuations and subject to Dynamical Coulomb Blockade in much the same way as a tunnel junction in the $P(E)$ theory,
- and one insensitive to voltage fluctuations, having consequently an intrinsic electrodynamic behavior like a conventional macroscopic electronic component.

Finally, the fluctuations of the electromagnetic environment modify not only the conductance of the SCC as is well known for instance for a tunnel junction, but it also reduces the quantum capacitance $-i\partial_\omega Y_{33}(\omega = 0)$ of the SCC, which is a new kind of DCB effect unveiled by our approach. In Sec. III. we discuss on an example how this effect on the quantum capacitance induces an electrostatic modification of the $\mathcal{S}$ matrix which may have large consequences on the transport properties of the system. Let us observe that, putting forward that in SCCs the Thouless energy $\hbar/\tau$ exceeds all other relevant energy scales, many works on SCCs with partially open channels simply disregard the energy-dependence of the $\mathcal{S}$ matrix and hence the quantum capacitance. Such approach consequently cannot apprehend the DCB of quantum capacitance effect we discuss here.

### E   Phase fluctuations

Now that we have determined how voltage fluctuations affect the current noise and admittance of the SCC through the correlators $e^{J_\pm(t)} = \langle e^{\pm i\varphi(t)} e^{\mp i\varphi(0)} \rangle$, it remains to determine these correlators in order to fully solve the transport problem in the circuit. To this end, we assume that the bulk effect of the fluctuations on transport can be captured satisfactorily by considering only second-order correlation functions of the currents and voltages (or, equivalently, phases), i.e. disregarding the non-Gaussianity of the fluctuations. This is our second key hypothesis. Within this approximation $J_\pm(t)$ express simply in terms of the phase correlator $S_{\varphi\varphi}(t) = \langle \varphi(t)\varphi(0) \rangle$ [5]

$$e^{J_\pm(t)} \simeq e^{S_{\varphi\varphi}(t) - S_{\varphi\varphi}(0)} \equiv e^{J(t)} \tag{10}$$

In essentially all the literature on Dynamical Coulomb Blockade, the phase autocorrelation function $S_{\varphi\varphi}(t)$ is evaluated using the Fluctuation-Dissipation Theorem (FDT)

$$S_{\varphi\varphi}(t) = 2\int_{-\infty}^{+\infty} \frac{d\omega}{\omega} \frac{\text{Re}Z_{\text{eff}}(\omega)}{R_K} \frac{e^{-i\omega t}}{1 - e^{-\beta\hbar\omega}} \tag{11}$$

with $Z_{\text{eff}}$ the effective environment impedance as seen from the SCC, and assumed in equilibrium at temperature $T$ (with $\beta = (k_B T)^{-1}$). Here, this expression is not applicable for two reasons. Firstly, since the SCC is described by an admittance matrix, the equivalent external circuit as seen from the SCC (which may contain other SCCs) needs to be described by a an impedance matrix $\mathbf{Z}_{\text{ext}}$ of the same size, or, equivalently, an admittance matrix $\mathbf{Y}_{\text{ext}} = \mathbf{Z}_{\text{ext}}^{-1}$ with a lumped element decomposition as in Fig. 1b. Secondly, the system we consider may be driven out of equilibrium by a dc voltage, in which case the FDT invoked in Eq. (11) is not applicable (nonequilibrium phase correlators have also been considered for tunnel junctions e.g. in Refs. [40, 41]).

Here we take a more general approach and consider the current noises of all parts of the circuit to act as sources which relax through all possible conduction paths, including the SCC, as shown in Fig. 1c. We then evaluate the resulting phase fluctuations in the circuit assuming that fluctuations are small enough so that the local linearity provided by the linear response theory adequately connects the current and phase fluctuations. This invocation of the linear response theory again involves only second-order correlators, in consistency with our approximation of Gaussian fluctuations. The matrix describing the phase fluctuations in the circuit

$$(S_{\varphi\varphi}(t))_{mn} = S_{\varphi_m \varphi_n}(t) = \langle \varphi_m(t) \varphi_n(0) \rangle \qquad m, n \in \{1, 3\}$$

is then linked to the current fluctuations through the (linear) matrix equation (for a derivation see the Supplemental Material)

$$S_{\varphi\varphi}(t) = \frac{1}{\hbar R_K} \int_{-\infty}^{+\infty} \frac{d\omega}{\omega^2} (Y(\omega) + Y_{\text{ext}}(\omega))^{-1} \cdot (S_{II}(\omega) + S_{II}^{\text{ext}}(\omega)) \cdot ((Y(\omega) + Y_{\text{ext}}(\omega))^{-1})^\dagger e^{-i\omega t} \qquad (12)$$

where $S_{II}(\omega) = \int S_{II}(t) e^{i\omega t} \frac{dt}{2\pi}$ is the current noise matrix of the SCC determined above in the frequency domain and $S_{II}^{\text{ext}}$ is the current noise matrix of the external circuit [42]. In such formulation, the SCC and its external surrounding circuit play a symmetric role in the determination of the phase fluctuations, thereby implementing the back-action of the SCC on the circuit. This extends what was done for large-conductance tunnel junctions in Ref. [43]. If the external circuit can be assumed at equilibrium, the FDT gives $S_{II}^{\text{ext}}(\omega) = 2 \text{Re}(Y_{\text{ext}}(\omega)) \frac{\hbar\omega}{1 - e^{-\beta\hbar\omega}}$. The two-point phase fluctuations across the SCC that enter the current fluctuations and admittance matrices (Eqs. (5), (8)) through Eq. (10), are given by

$$S_{\varphi\varphi}(t) = [1 \ -1 \ 0] \cdot S_{\varphi\varphi}(t) \cdot \begin{bmatrix} 1 \\ -1 \\ 0 \end{bmatrix} \qquad (13)$$

while the phase correlator of the internal node, responsible for the "noise at the gate" [44, 39] is $(S_{\varphi\varphi}(t))_{33}$. In a tunnel junction the phase correlator Eq. (11) is indeed recovered from Eqs. (12) and (13) if one assumes that the current in the junction is so small that it has no back-action on the rest of the circuit which hence remains in equilibrium: then one has $|S_{II}(\omega)| \ll |S_{II}^{\text{ext}}(\omega)|$ (implying $|Y(\omega)| \ll |Y_{\text{ext}}(\omega)|$), $S_{I_2 I_2} = S_{I_1 I_1} = -S_{I_1 I_2}$ (forbidding any internal degree of freedom), and equilibrium noise. Hence, our expressions Eqs. (12), (13) broadly generalize the evaluation of the phase correlator made in the standard $P(E)$ theory to components with an internal degree of freedom, that have an arbitrarily large back-action on the circuit and to out-of-equilibrium situations. Furthermore, this formulation is directly usable for multiterminal SCCs as well. Given that it ensures the global consistency of fluctuations in the circuit in a systematic manner independently of any detail of the circuit, we think it constitutes another important step taken in this work.

### F  Closing the loop

In our approach, $S_{\varphi\varphi}$, $S_{II}$ and $Y$ are obviously inter-dependent quantities so that the set of equations (5), (10), (12) and (13) needs to be solved self-consistently. Such a self-consistent approach for the electrodynamics was used previously to describe low-resistance normal-state tunnel junctions [43] and Josephson junctions [45]. In the former case, it has been successfully checked experimentally, and it was shown to correspond to a self-consistent harmonic approximation that minimizes the free energy in the path integral description of the system [46]. The self-consistent solution discussed here much extends previous results as it can now handle arbitrarily strong non-linear interaction between an SCC of arbitrary transmission with any (possibly nonlinear) surrounding circuit, including in out-of-equilibrium situations, provided the non-Gaussianity of fluctuations can be disregarded. Yet, the nonlinearity is still captured owing to the factorization of the phase correlation function $e^{J(t)}$ in part of the current noise (Eq. (5)). In the frequency domain (e.g. in $Y(\omega)$), this factorization formally becomes a convolution product (with the $P(E)$ function [5], precisely) that mixes different frequencies, as expected in a nonlinear system.

At this point we can sketch how the interaction operates : The partition noise of quasiparticles at the SCC generates plasmonic modes whose current and voltage fluctuations are related by the (locally linear) electrodynamics of the whole circuit. The low-frequency, large wavelength plasmonic (bosonic) modes propagating in the conductors mediate an effective electron-electron interaction among branches in the whole circuit (and even among the partially open Landauer channels of the SCC itself), corresponding to a Random Phase Approximation (RPA) treatment of electron-electron interactions. This effective interaction is normally ignored in LB approach build on strictly non-interacting Landau quasiparticles; its effect is here encapsulated in the $e^{J_\pm(t)}$ terms of Eq. (5). In several theoretical works, this coupling of quasiparticles with the electromagnetic field is handled through the "bosonization" technique [11, 17]. In the tunnel limit the link between the electron-electron interaction and the phase correlator $e^{J_\pm(t)}$ was discussed in [47, 48, 5].

### G  Validity of Approximations

Let us now recap and discuss the two key hypothesis we made: (i) the decoupling of timescales between electronic scattering and electromagnetic fluctuations, and (ii) phase fluctuations can be regarded as being Gaussian in order to capture the bulk of the transport properties.

(i) Our hypothesis of rapid quasiparticle scattering compared to the typical timescale of electromagnetic fluctuations will be consistent if $\exp J(\tau) \sim 1$, meaning that during the time an electron crosses the SCC, it sees essentially a static field configuration, and different configurations of the fluctuating voltage are averaged in the succession of quasiparticle scattering events. Otherwise, as already said, $\tilde{S}$ cannot be simply obtained from $S$ defined in absence of an environment as we have assumed.

(ii) Gaussian fluctuations is the bare minimum one may consider when aiming for a theory incorporating fluctuations. Electronic circuits have rigorously Gaussian fluctuations only when the electrodynamics is purely linear and the system is in a thermal coherent state [49]. Here, the system is clearly non-linear and it can moreover be kept out-of-equilibrium by a dc voltage, so that the range of validity of our approximation needs to be checked. Putting theoretical boundaries to this approximation would require to consider higher order cumulants [50, 51], which is beyond the scope of this paper. Nevertheless, we qualitatively expect our approach to fail in some cases, for instance when phase fluctuations become large in systems where charge quantization effects in some part of the circuit cannot be ignored.

## III.  WORKING OUT A CONCRETE CASE: THE QUANTUM POINT CONTACT

In the remainder of the article we illustrate how our theory operates in the case where the SCC is a single-channel Quantum Point Contact (QPC) in series with an R-C impedance (Fig. 2a). For such a setup, measurements of the reduction of the conductance due to an on-chip R-C environment (i.e. DCB effect) were carried out by Pierre and coworkers [9, 52, 18]. The results of such measurements are shown in Fig. 3 as a function of the "intrinsic" transmission of the QPC. Let us recall that these experiments where shown to follow scaling laws [52] predicted by the mapping of the system to an impurity in a TLL [11, 13, 18] and by a renormalization group approach [12, 10], but these scalings require a reference point which they cannot predict.

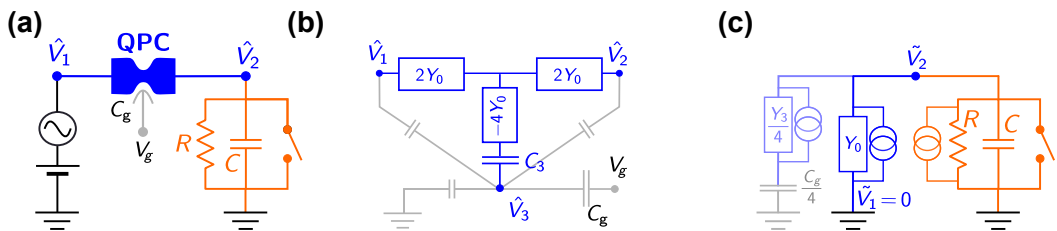

FIG. 2. (a) Quantum Point Contact (QPC) in series with an R-C circuit. The circuit can be biased by a dc voltage source and an additional small ac excitation enables measuring its differential conductance using a lock-in technique. In the experimental implementation the R-C circuit could be short-circuited on-chip in order to measure the effect of its presence. When the switch is closed, the QPC is voltage biased: voltage fluctuations are suppressed and the usual LB description applies. (b) Lumped element model of the QPC. Assuming the QPC is symmetric w.r.t. exchange of nodes 1 and 2, the lumped element model of Fig. 1b has only two independent elements and can be represented as shown here, where $C_3(\omega) = Y_3(\omega)/i\omega$ is the quantum capacitance of the conductor, and $Y_0(\omega)$ its 2-point admittance. In this panel we also symbolically represent the internal node of the device and, in grey, the geometrical capacitances that may affect it and which are not (and cannot be) included in the admittance matrix obtained from the electronic scattering matrix. (c) Small-signal equivalent of the full circuit used to determine voltage fluctuations across the QPC. In the simple case we consider here, the general analysis (see Eq. (12) and Fig. 1c) reduces to a simple parallel combination of 2-terminal components. $Y_3$ is greyed out because it has a negligible contribution in the fluctuations across the QPC.

### A first simplifications

Compared to the formal general case discussed in sec. II., this experimental setup brings several simplifications: (i) the QPC is tuned to have a single channel (with the spin degeneracy lifted by a magnetic field) so that the $\mathcal{S}$ matrix is only $2 \times 2$, (ii) due to the symmetric design of the QPC, its $\mathcal{S}$ matrix can be assumed symmetric w.r.t. leads 1 and 2, and (iii) it was experimentally checked that heating effects were negligible, enabling us to use Fermi functions at the experiment's temperature for the electronic distribution functions in both reservoirs. The correlators have the symetries $S_{I_2 I_2}(V) = S_{I_1 I_1}(-V) = S_{I_1 I_1}(+V)$ w.r.t. $V = V_1 - V_2$ the dc voltage drop across the conductor, so that $Y_{22} = Y_{11}$. Hence, only two independent noises or admittances are needed to describe the QPC. In that case it is convenient to describe transport in terms of the common mode current $I_3$ already introduced above and the differential mode current $I_0 = \frac{1}{2}(I_1 - I_2)$. The fluctuations $S_{I_0 I_0}$ and $S_{I_3 I_3}$ as well as the corresponding admittances $Y_0$ and $Y_3$ are readily obtained from the elements of $\boldsymbol{S}_{II}(t)$ and $\boldsymbol{Y}(\omega)$ in sec. II.:

$$
\begin{align}
S_{I_0 I_0} &= \frac{1}{2}(S_{I_1 I_1} - S_{I_1 I_2}) \tag{14}\\
S_{I_3 I_3} &= 2(S_{I_1 I_1} + S_{I_1 I_2}) \tag{15}\\
Y_0 &= \frac{1}{2}(Y_{11} - Y_{12}) \tag{16}\\
Y_3 &= 2(Y_{11} + Y_{12}) \tag{17}
\end{align}
$$

all these quantities implicitly depending on the dc bias voltage $V$. The admittance $Y_0$ describes transport through the QPC when, in addition to $V$, a small symmetric differential ac voltage is applied on the QPC; it yields the usual 2-terminal differential conductance in the zero-frequency limit, $\frac{dI(V)}{dV} = Y_0(\omega = 0)$ and subsequently the $I - V$ characteristics by integration over $V$. On the contrary, $Y_3$ gives the response of the QPC submitted to the same small ac voltage on both leads; it describes charge accumulation in the QPC and is essentially capacitive at low frequency ($Y_3(\omega \sim 0) \sim iC_3 \omega$), with $C_3 = e^2 D$, $D = dN/dE$ being the local density of states at the QPC [24, 27]. Fig. 2b shows a lumped model description of the QPC. This model can be seen as the result of a triangle-star (a.k.a. $\Delta$-Y) circuit transformation of Fig. 1b, specialized to the case we consider here. In Fig. 2b we also show additional geometrical capacitances which are not taken into account in the admittance matrix $\boldsymbol{Y}$ since it derives from the scattering matrix that describes only quasiparticle transfers and not displacement currents. Finally, in this case the general small-signal linear response analysis leading to the phase correlation function (Eqs. (12) and (13)) is simplified (see Fig. 2c vs Fig. 1c for the general case):

$$
S_{\varphi\varphi}(t) = \frac{2}{R_K} \int_{-\infty}^{+\infty} \frac{d\omega}{\omega^2} \left( S_{I_0 I_0}(\omega) + \frac{1}{4} \frac{iC_g \omega}{Y_3(\omega) + iC_g \omega} S_{I_3 I_3}(\omega) + \right. \tag{18}
$$

$$
\left. \frac{2R}{1+(RC\omega)^2} \frac{\hbar\omega}{1-e^{-\beta\hbar\omega}} \right) |Z_{\text{eff}}(\omega)|^2 e^{-i\omega t}
$$

where

$$
Z_{\text{eff}}(\omega) = \left( Y_0(\omega) + \frac{1}{4} \frac{iC_g \omega Y_3(\omega)}{Y_3(\omega) + iC_g \omega} + \frac{1}{R} + iC\omega \right)^{-1} \tag{19}
$$

and which further reduces to Eq. (11) in equilibrium.

### B Model for the scattering matrix

The general form of the quasiparticle scattering matrix for the time-reversal symmetric and spatially symmetric single-channel QPC is [53, 54]

$$
\mathcal{S}(E) = e^{i\phi(E)} \begin{pmatrix} -i r(E) e^{i\alpha} & t(E) \\ t(E) & -i r(E) e^{-i\alpha} \end{pmatrix} \tag{20}
$$

in which $\alpha$ is a real number and $\phi(E)$, $t(E)$, $r(E)$ are all real functions, with $r(E)^2 + t(E)^2 = 1$ ensuring particle conservation. The usual way to model a QPC is to assume a parabolic saddle-like potential at the constriction which, for the first mode, leads to a transmission probability known as the Kemble formula [55, 22], similar to a Fermi function

$$\mathcal{T}(E) = t(E)^2 = \frac{1}{e^{2\pi[E_0 - E]/\hbar\Omega} + 1} \quad (21)$$

with $E_0$ being the maximum energy of the effective 1D potential for the first mode at the constriction and $\Omega$ the characteristic frequency of the electrons in this 1D parabolic potential. In this model, both the global phase $\phi(E)$ and the (constant) backscattering phase $\alpha$ drop out of this probability and other results. What is often simply called the "channel transmission" in e.g. the Landauer formula is actually the transmission taken at the Fermi level $\mathcal{T}_0 \equiv \mathcal{T}(E = 0)$. There is a biunivocal relation between the saddle point energy $E_0$ and $\mathcal{T}_0$:

$$E_0 = \frac{\hbar\Omega}{2\pi} \log\frac{1 - \mathcal{T}_0}{\mathcal{T}_0} \quad (22)$$

and $\mathcal{T}_0$ can itself be obtained from the conductance at temperatures $T \ll \hbar\Omega/k_B$ and in absence of voltage fluctuations. In experiments, when the gate voltage is changed, the electronic fluid adapts to the externally applied field resulting in a new self-consistent effective potential barrier height. Likewise, for a finite bias voltage across the barrier, its shape is expected to change following a redistribution of charges [56]. Here we simply assume the shape of the barrier can be regarded as fixed in the bias voltage range used in the experiment. In that case the effect of a dc voltage between the reservoirs can be simply accounted for by including it in the scattering phase $\varphi(t) \to \varphi(t) + ieVt/\hbar$ in $\tilde{S}$ (Eq. (4)), leading to phase correlators $\langle e^{\pm i\varphi(t)} e^{\mp i\varphi(0)}\rangle_{\text{em}} = e^{J(t)} e^{\pm ieVt/\hbar}$.

Values for $\hbar\Omega$ depend on the exact geometry of the QPC, and for short QPC they are typically a fraction of meV.

### C  Further approximations

At this point we could proceed and evaluate numerically Eqs. (5), (8), (16-18) until self-consistence is reached. However such numerical solution would essentially hide how the theory actually operates in detail, thus making it harder to reach a thorough understanding. In order to better expose how our theory works, in the following we rather recourse to analytical methods. To this end, we exploit the initial hypothesis that $\Omega$ is the fastest dynamics in the system which enables further approximations. This approach also yields new analytical results that can be related to previous works.

#### 1.  Energy-independent transmission

Let us first consider the limit where $\Omega \to \infty$. In that limit the energy dependence of scattering matrix of the QPC can be disregarded, making the replacement $\mathcal{T}(E) \to \mathcal{T}_0$. Then Eqs. (5), (14) and (15) give $S_{I_3 I_3} = 0$ (there is no internal degree of freedom in this limit) while

$$S_{I_0 I_0}(t) = 2\pi\frac{G_K}{\hbar}\left((1-\mathcal{T}_0)\mathcal{T}_0 e^{J(t)} \gamma(t) 2\cos\frac{eVt}{\hbar} + \mathcal{T}_0^2 2\gamma(t)\right) \quad (23)$$

where $\gamma(\epsilon) = \int d\epsilon' f(\epsilon')(1-f(\epsilon'+\epsilon)) = \frac{\epsilon}{1-e^{-\epsilon/k_B T}}$, $\gamma(t) = \int d\epsilon\, \gamma(\epsilon)\, e^{-i\epsilon t/\hbar}$ its inverse Fourier transform, $f$ being the Fermi function and $T$ the temperature of the reservoirs. In the spectral domain $S_{I_0 I_0}(\omega) = \int \frac{dt}{2\pi} S_{I_0 I_0}(t) e^{i\omega t}$, using $P(E) = \frac{1}{2\pi\hbar}\int_{-\infty}^{\infty} e^{J(t) + iEt/\hbar} dt$ this becomes

$$S_{I_0 I_0}(\omega) = G_K((1-\mathcal{T}_0)\mathcal{T}_0(\gamma * P(\hbar\omega + eV) + \gamma * P(\hbar\omega - eV)) + \mathcal{T}_0^2 2\gamma(\hbar\omega))$$

where $*$ denotes convolution. In the above expressions for the current noise, the $(1-\mathcal{T}_0)\mathcal{T}_0$ term is identical to the current noise of a tunnel junction [20] with tunnel conductance $G_K(1-\mathcal{T}_0)\mathcal{T}_0$, while the $\mathcal{T}_0^2$ term also corresponds to a tunnel junction but at zero voltage, without environment ($e^{J(t)} = 1$; $P(E) = \delta(E)$) and a conductance $G_K \mathcal{T}_0^2$, the later junction being hardly distinguishable from a macroscopic resistor. This decomposition was already obtained in Refs. [7, 6] in the case of weak DCB.

Applying the linear response formula Eq. (8) we immediately obtain $Y_3 = 0$, and, using known results for the DCB of a tunnel junction in real time formulation [43], the admittance of the QPC

$$Y_0(\omega, V) = G_K \mathcal{T}_0 [1 + (1 - \mathcal{T}_0) F(\omega, J, V, T)], \tag{24}$$

where

$$F(\omega, J, V, T) = \frac{2\pi}{(\hbar\beta)^2} \int_0^{+\infty} \left(\sinh\frac{\pi t}{\hbar\beta}\right)^{-2} \operatorname{Im} e^{J(t)} \cos\frac{eVt}{\hbar} \frac{e^{-i\omega t} - 1}{-i\omega} dt \tag{25}$$

(for the $\mathcal{T}_0^2$ noise term, there is no $F$ term). In that limit the QPC behaves as a genuine 2-terminal conductor and its dynamics is that obtained for a resistor in parallel with a high conductance tunnel junctions in [43], with, in particular, a self-shunting effect taken into account in $J(t)$. It also generalizes to arbitrary impedances the $(1 - \mathcal{T}_0)$ linear suppression of Dynamical Coulomb Blockade previously derived only in the small-impedance, weak blockade limit [6, 7] and observed in experiments [8, 9].

In this $\Omega \to \infty$ limit one may drop both $S_{I_3 I_3}$ in Eq. (18) and $Y_3$ (or $C_3$) in Eq. (19) (they are both $\mathcal{O}(\Omega^{-1})$). This yields

$$J(t) = \frac{2}{R_K} \int_{-\infty}^{+\infty} \frac{d\omega}{\omega^2} \left( S_{I_0 I_0}(\omega) + \frac{2R}{1 + (RC\omega)^2} \frac{\hbar\omega}{1 - e^{-\beta\hbar\omega}} \right) \left| \frac{1}{Y_0(\omega) + \frac{1}{R} + iC\omega} \right|^2 (e^{-i\omega t} - 1) \tag{26}$$

which in equilibrium (at $|eV| \lesssim k_B T$) reduces to

$$J(t) = \frac{2}{R_K} \int_{-\infty}^{+\infty} \frac{d\omega}{\omega} \operatorname{Re}\left[\frac{1}{Y_0(\omega) + \frac{1}{R} + iC\omega}\right] \frac{e^{-i\omega t} - 1}{1 - e^{-\beta\hbar\omega}}. \tag{27}$$

The actual phase correlation function $J$ is obtained from the self-consistent solution of Eqs. (23), (24) and (26) (or (27) in equilibrium); it depends on the total effective impedance as seen from the QPC leads, itself depending through $Y_0$ on the transmission $\mathcal{T}_0$ of the QPC, on the temperature and on the dc voltage. In the absence of environment $R \to 0$, so that $J(t) = 0$ and one recovers the usual LB results, not requiring self-consitency. Solving numerically the self-consistent equations in this energy-independent transmission limit in presence of the impedance, we predict that the relative reduction of the conductance (w.r.t. the Landauer conductance) is nearly linear with the transmission of the QPC (see top right panel of Fig. 3), with a tiny convexity due to the self-shunting effect of $Y_0$ in $J(t)$. Obviously this does not account for the experimental data of Pierre and co-workers, and one needs to consider the energy-dependence of the scattering matrix, at least in leading order.

In any case, the function $F(\omega, J, V, T)$ entering the admittance is bounded between 0 and $-1$, with the actual value depending on the relative value of the energy scales $k_B T$, $eV$, $\hbar\omega$ and $\hbar\omega_c$, where $\omega_c$ is the cutoff frequency of the total admittance $Y_0(\omega) + \frac{1}{R} + iC\omega$ determining the phase correlator. When $\operatorname{Max}(k_B T, eV) \gg \hbar\omega_c$, $F \sim 0$ and DCB is suppressed, while for $\operatorname{Min}(k_B T, eV, \hbar\omega) \ll \hbar\omega_c$, $F \sim -1$ corresponding to maximum strength DCB. Thus, the dc conductance $Y_0(\omega = 0, V)$ of the QPC is such that $\mathcal{T}_0^2 \leqslant R_K Y_0(\omega = 0, V) \leqslant \mathcal{T}_0$ showing that, as already anounced, the usual "Landauer formula" does not hold in the DCB regime.

*2. Energy-dependent transmission: leading order terms*

The leading order term in $Y_3$ that enables capturing the effect of the environment on the internal degree of freedom for large but non infinite $\Omega$ is $\mathcal{O}(\Omega^{-1})$. We evaluate it by taking the zero-temperature limit $S^0_{I_3 I_3}$ of $S_{I_3 I_3}$ (Eqs. (15),(5)), but with the phase correlations still evaluated at finite temperature and voltage. Hence one simply replaces the distribution functions by Heaviside step functions

$$f_{\pm i}(\varepsilon) \to \theta(\mp\varepsilon),$$

in which case the energy integrals in Eq. (5) can all be evaluated analytically. The corresponding analytical result for $S^0_{I_3 I_3}$ writes as

$$S^0_{I_3 I_3} = S^{0R}_{I_3 I_3}(t, E_0) + \cos\frac{eVt}{\hbar} e^{J(t)} S^{0J}_{I_3 I_3}(t, E_0)$$

where $S^{0R}_{I_3I_3}(t,E_0)$ and $S^{0J}_{I_3I_3}(t,E_0)$ are respectively the "resistor-like" and "junction-like" noise of the common mode current $I_3$, whose large expression involving special functions are given in the Supplemental Material. In the following we actually need only the zero-frequency quantum capacitance

$$\begin{aligned}C_3(E_0,J,V) &= -i\partial_\omega Y_3(\omega=0) \\ C_3(E_0,J,V) &= \frac{1}{\hbar}\int_0^\infty t^2 \mathrm{Im}\left[S^{0R}_{I_3I_3}(t,E_0)+\cos\frac{eVt}{\hbar}e^{J(t)}S^{0J}_{I_3I_3}(t,E_0)\right]\frac{dt}{2\pi}\end{aligned} \quad (28)$$

which depends on the channel transmission $\mathcal{T}_0$ (or equivalently, $E_0$), the phase correlation function $J(t)$ and the dc voltage. Using the fact that $S_{I_3I_3}(t)$ is peaked at short times ($t \lesssim \Omega^{-1}$), and the hypothesis that the phase does not vary much on that time scale, it is then justified to expand $e^{J(t)} \simeq 1 + J'(0)t$ in Eq. (28), and furthermore using [5]

$$\mathrm{Im}[J'(0)] = -\frac{\pi}{R_K C_{\mathrm{tot}}},$$

with $C_{\mathrm{tot}}$ being the total capacitance as seen from the QPC (including possibly $-i\partial_\omega Y_0(\omega=0)$, see Eq. (27)), one can evaluate the DCB correction to the quantum capacitance

$$C_3(E_0,J,V) \simeq C_3(E_0,J=0,V) - \frac{\pi}{\hbar R_K C_{\mathrm{tot}}}\int_0^\infty t^3 \cos\frac{eVt}{\hbar}\mathrm{Re}[S^{0J}_{I_3I_3}(t,E_0,J=0)]\frac{dt}{2\pi}. \quad (29)$$

Interestingly, in this last expression the environmental resistance $R$ drops out (but the timescale decoupling hypothesis still requires $RC_{\mathrm{tot}}\Omega \gg 1$). This expression also confirms that in our approach the quantum capacitance is modified by the electromagnetic environment and differs from that predicted in the usual LB approach. Our result Eq. (29) generalizes to quantum fluctuations the change of the quantum capacitance produced by a classical ac drive [30] that also only occurs if the scattering matrix is energy-dependent.

At the $\mathcal{O}(\Omega^{-1})$ order we consider here, the approximations leading to the phase correlator Eq. (26) (or (27) in equilibrium) are still valid. Thus, $J(t)$ does not depend directly on $C_3$, which hence does not enter in the self-consistent determination of the admittance at a given transmission. On the other hand $C_3$ depends on the channel transmission both directly through the expression of $S^0_{I_3I_3}$ and indirectly through the $\mathcal{T}_0$-dependence of $Y_0(\omega)$ in $J$. Below we discuss how $\mathcal{T}_0$ depends on $C_3$ through electrostatics, thereby creating another level of self-consistency.

### D  DCB of the quantum capacitance: modification of the transmission

Consider the though experiment where, by adjusting the gate voltage, one sets the *bare* transmission $\mathcal{T}_0$ of the QPC in absence of any external impedance (in that case the usual Landauer formula applies: the conductance is a measurement of $\mathcal{T}_0$) and subsequently inserts the $RC$ impedance in series with the QPC. As discussed in general terms above, the phase fluctuations due to the $RC$ environment partly reduce both $Y_0$ and $Y_3$ w.r.t. the "bare" case, reducing respectively the conductance (see Fig. 3) and the quantum capacitance $C_3$ of the QPC. The reduction of $C_3$ (i.e. the LDOS) changes the dc charge configuration in the QPC which impacts the self-consistent effective 1D potential and notably its height. Consequently the transmission $\mathcal{T}_0$ of the channel set in absence of environment takes a new value $\mathcal{T}_0^*$. In models taking into account electron-electron interactions in a QPC [57], a similar reduction of the LDOS is obtained by increasing the interaction parameter, resulting in a reduction of the conductance at a fixed bare barrier height. A reduction of the LDOS is also predicted in a TLL with an impurity [58] which is known to map onto the QPC with ohmic environment considered here [11]. In our approach interactions are handled by the electrostatics of the system like in Refs [54, 59].

Let us consider the electrostatic configuration of the channel, including the effect of the geometrical capacitances depicted in grey in Fig. 2b. In the experiment the gate is made of metal with a much larger density of states than the 2-DEG of the QPC so that it has a negligible quantum capacitance [29]. The difference of the electron charge $eN(E_0)$ and the background charge $eN^+$ in the channel is equal to the total charge on the geometrical capacitors from the channel to the other conductors [60]

$$eN(E_0) - eN^+ = C_g(V_3 - V_g) + C_1(V_3 - V_1) + C_2(V_3 - V_2) \quad (30)$$

where $C_1$ and $C_2$ denote the geometrical capacitances from the channel to the reservoirs (further geometrical capacitance may be added as needed). Furthermore, the energy barrier height is simply $E_0 = -eV_3$, and the total number of electrons in the QPC for a given barrier height is

$$N(E_0) = \frac{1}{e^2} \int_{E_0}^{E_{\max}} C_3(E) dE \tag{31}$$

with the LDOS given by the quantum capacitance

$$\frac{1}{e^2} C_3 = \frac{dN}{dE}(E_F) = -\frac{dN}{dE_0}$$

the last equality coming from the assumption that the shape of the potential is fixed and controlled by the barrier height. In Eq. (31), $E_{\max}$ is a cut-off energy where the channel is assumed fully depleted ($E_{\max} \gg \hbar\Omega$) and which is needed to avoid a logarithmic divergence due to the naive QPC model we use. Eq. (30) then links the barrier height to the various voltages sources in the circuit

$$C_{\text{gtot}} E_0 + \int_{E_0}^{E_{\max}} C_3(E) dE = -eC_g V_g - eC_1 V_1 - eC_2 V_2 + eN^+ \tag{32}$$

with $C_{\text{gtot}}$ being the sum of all geometrical capacitance connecting the channel to other nodes.

We now consider the two situations of the thought experiment where the QPC sees a zero (respectively finite RC) impedance environment, and we denote its barrier height $E_0$ (resp. $E_0^*$), and its LDOS is given by $C_3(E_0, J = 0)$ (resp. $C_3^* = C_3(E_0^*, J(E_0^*))$) where we made explicit that, according to Eq. (28), $C_3$ depends on the barrier height also through the phase correlator $J$. By considering Eq. (32) in the two different situations for the same gate and reservoir voltages one may relate $E_0$ and $E_0^*$

$$C_{\text{gtot}} E_0^* + \int_{E_0^*}^{E_{\max}} C_3^*(E) dE = C_{\text{gtot}} E_0 + \int_{E_0}^{E_{\max}} C_3(E) dE$$

which leads to the differential relation linking $E_0$ and $E_0^*$:

$$\frac{dE_0^*}{dE_0} = \frac{C_{\text{gtot}} - C_3(E_0)}{C_{\text{gtot}} - C_3^*(E_0^*)} \tag{33}$$

where $C_3(E_0, J=0)$ is known analytically from the zero temperature result (Eq. (28), with $e^{J(t)} = 1$, and results in the Supplemental Material), and $E_0^*(E_0)$ is the value of the DCB-affected energy barrier as a function of the bare energy barrier. The initial condition for this differential equation is $E_0^*(E_0 = E_{\max}) = E_{\max}$ where it is assumed that no electrons are left in the channel so that they cannot alter the potential. Using Eqs. (22) and (21) one then obtains the DCB-modified transmission $\mathcal{T}_0^*(\mathcal{T}_0) = \mathcal{T}(E_0^*(E_0(\mathcal{T}_0)))$ and we predict the DCB-modified admittance of the QPC is given by Eqs. (24), (25) with $\mathcal{T}_0^*$ in place of $\mathcal{T}_0$:

$$Y_0(\omega, V) = G_K \mathcal{T}_0^* \left[ 1 + (1 - \mathcal{T}_0^*) \frac{2\pi}{(\hbar\beta)^2} \int_0^{+\infty} \left( \sinh \frac{\pi t}{\hbar\beta} \right)^{-2} \operatorname{Im} e^{J(t)} \cos \frac{eVt}{\hbar} \frac{e^{-i\omega t} - 1}{-i\omega} dt \right]. \tag{34}$$

This strong prediction is the central result of this part and it stresses the distinctions to be made between the bare transmission $\mathcal{T}_0$, the DCB-modified transmission $\mathcal{T}_0^*$ and the dimensionless dc conductance $\mathcal{T}_0^{*2} \leqslant R_K Y(\omega = 0, V = 0) \leqslant \mathcal{T}_0^*$ in the DCB regime. Equation (33) furthermore shows that the change of the transmission upon inserting the impedance is not universal: The modified transmission inherits the $C_3$-dependence on the bias voltage and on the temperature which depends on the type of scatterer considered and, even for a given type of scatterer, the relative value of the quantum capacitance and the total geometric capacitance of the channel of the QPC depend on the sample geometry, the dielectric constants, the 2DEG density, etc.

The experiments of Pierre and coworkers tried to closely implement this thought experiment: The samples incorporated an on-chip switch that enabled to short-circuit the resistance (see Fig. 2a), suppressing the fluctuations of the environment. When the switch is closed, one hence measures a zero-bias conductance given by the bare transmission $\mathcal{T}_0$ of the channel, and, when the switch is opened, a DCB-reduced conductance. However, the switch being itself an auxiliary QPC, switching it without changing the setpoint of the primary QPC requires a careful cancellation of the capacitive crosstalk between the QPCs. This cancellation relies on measurements performed on the system tuned at different operation points and there is no means of ensuring directly that the cross-talk is properly canceled while the DCB measurement are taken, leaving the possibility that a small systematic error is made. In order to avoid this possible crosstalk cancellation issue, Pierre and coworkers have used another protocol for some data sets shown in Fig. 3 (see supplementary material of Ref. [18] and Table I): they measured the conductance in presence of the environmental impedance but at a bias voltage $eV_\infty \gg \hbar/RC$ where the voltage dependence flattens outs, extracting a transmission $\mathcal{T}_\infty$ which they assumed somehow identical to the bare transmission (checking this assumption itself brings back to the crosstalk cancellation issue). In our analysis however, the quantum capacitance in the presence of the impedance clearly differs (at all voltages) from its bare value because of the second rhs term in Eq. (29). Consequently, from the electrostatic equilibrium standpoint, such a "high voltage" measurement is not equivalent to shunting the environment and consequently does not give access to the bare channel transmission $\mathcal{T}_0$. Using our analysis of the channel's electrostatics one may still relate the "high voltage" energy barrier $E_\infty$ (corresponding to transmission $\mathcal{T}_\infty$) to the DCB-affected zero-voltage energy barrier $E_0^*$ through the differential equation

$$\frac{dE_0^*}{dE_\infty} = \frac{C_{\text{gtot}} - C_3^*(E_\infty, V_\infty)}{C_{\text{gtot}} - C_3^*(E_0^*, V=0)}. \tag{35}$$

However, in our simple model the total channel charge $N(E_0)$ (Eq. (31)) entering the electrostatic balance Eq. (30) (that yields Eq. (35)) is such that its impedance-independent part depends differently on the high energy cutoff $E_{\max}$ at zero and finite voltage. Therefore, unlike for Eq. (33) where all quantities are taken at the same (null) voltage, setting the initial condition for Eq. (35) is not independent of the choice of the upper energy bound. Within our model this cutoff dependance can be solved by the introduction of an additional adjustable parameter $U$ in the initial condition for Eq. (35) $E_0^*(E_\infty = E_{\max}) = E_{\max} + U$ w.r.t. the fully zero bias protocol using the switch.

### E   Comparison with the experimental results of Pierre and co-workers

We now compare the results of the previous section to the experimental results of Pierre and coworkers. The procedure used for the data set with $R = 6.3$ k$\Omega$ which was measured using the switch protocol is the following. One first solves self-consistently transport in the QPC with a given fixed transmission (i.e. a given $E_0$) of the QPC, in presence of the environment: For the given value of $E_0$, we define a proper sampling of the functions $J(t)$ and $Y_0(\omega)$ and initialize $Y_0(\omega) = G_K \mathcal{T}_0$. Then we iterate numerically Eqs. (23), (24) and (27) until suitably converged. We repeat this calculation for a set of energies $-E_{\max} \leq E_0 \leq E_{\max}$ with $E_{\max} \gg \hbar\Omega$, i.e. exploring the whole range of transmissions $\mathcal{T}_0 \sim 0 \to \mathcal{T}_0 \sim 1$. This yields tabulated values of the conductance $G(E_0) = Y_0(\omega=0, E_0)$ and of $C_3^*(E_0)$. In a second step, using the tabulated values of $C_3^*$ we numerically solve the differential equation Eq. (33) with the initial condition $E_0^*(E_{\max}) = E_{\max}$. While doing this, the ratio of the geometrical capacitance to the quantum capacitance (more precisely, $C_{\text{gtot}} \hbar\Omega/e^2$) is our single adjustable parameter. Let us stress that this parameter could in principle be evaluated from the QPC geometry and material parameters, thereby making our theory fully predictive. Finally, using this solution and the tabulated conductance, one predicts the Coulomb-blockaded QPC admittance in the experiment, for a given transmission $\mathcal{T}_0$ in absence of environment as $G(E_0^*(E_0(\mathcal{T}_0)))$. Using Eq. (21), we can as well plot the Coulomb-blockaded channel transmission $\mathcal{T}_0^*$ as a function of the bare (or high voltage) channel transmission as $\mathcal{T}_0^*(\mathcal{T}_0) = \mathcal{T}(E_0^*(E_0(\mathcal{T}_0)))$ (see bottom right panel of Fig. 3). For the datasets $R \in \{R_K/2, R_K/3, 80 \text{ k}\Omega\}$) which were measured using the "high voltage" protocol yielding the $\mathcal{T}_\infty$ reference transmission, the procedure is similar, but requires and additional set of self-consistent calculations at bias voltage $V_\infty$ and using the differential equation Eq. (35) with the initial condition $E_0^*(E_{\max}) = E_{\max} + U$, with $C_{\text{gtot}}$ and $U$ as adjustable parameters.

In Fig. 3 we compare the predictions of our approach to the experimental data of Ref. [18]. When doing so, we use the values of the environmental resistance and of the capacitances that are given in Ref. [18]. As for the temperatures, we have also used the values given in Ref. [18] for the 6.3 k$\Omega$ and 80 k$\Omega$ samples, but for the $R_K/2$ and $R_K/3$ data (that were taken on the same sample see Ref. [18]) we have used $T = 22$ mK [61], instead of the reported 16 mK and 17 mK. We found this change necessary for the model to recover the quite precisely measured conductance in the tunnel limit for the $R_K/3$ data, and because not adjusting this starting value would spoil the adjustment at all other transmissions. The values of the adjustable parameters used to produce the theoretical curves in Fig. 3 are given in Table I.

| $R$ | 6.3 kΩ | $R_K/3$ | $R_K/2$ | 80 kΩ |
|---|---|---|---|---|
| $C$ | 2.3 fF | 2.2 fF | | 2.8 fF |
| $T$ | 54 mK | (17 mK) 22 mK | (16 mK) 22 mK | 20 mK |
| $(\delta G/G_K \mathcal{T}_{\text{ref}})_{\mathcal{T}_{\text{ref}} \to 0}$ | −62% | (−85%) −82% | (−92%) −89% | −98% |
| $\mathcal{T}_{\text{ref}}$ (exp. protocol) | $\mathcal{T}_0$(switch) | $\mathcal{T}_\infty$ ($V_\infty = 63\,\mu$V) | | $\mathcal{T}_\infty$ ($V_\infty = 82\,\mu$V) |
| $2\pi U/\hbar\Omega$ | – | 0.45 | 0.7 | 1.75 (*) |
| $C_{\text{gtot}}/\text{Max}(C_3)$ | 1.37 | 1.36 | | 2.5 (*) |

TABLE. I Sample parameters and adjustable parameters used in Fig. 3. The sample parameters $R$, $C$ and $T$ were taken from Ref. [18]. The temperature value for $R = R_K/3, R_K/2$ which the authors gave (in parenthesis) was however raised to 22 mK in order for the conductance reduction predicted in the tunnel limit by the $P(E)$ theory (third row) to better agree with the measured conductance reduction. The fourth row indicates which experimental protocol (see text) was used to provide the reference transmission $\mathcal{T}_{\text{ref}}$. The last two rows are the adjustable parameters (see text) used to provide the adjustments shown in Fig. 3. For the $R = 80$ kΩ sample, the parameters indicated by an asterisk can be raised simultaneously without affecting much the prediction.

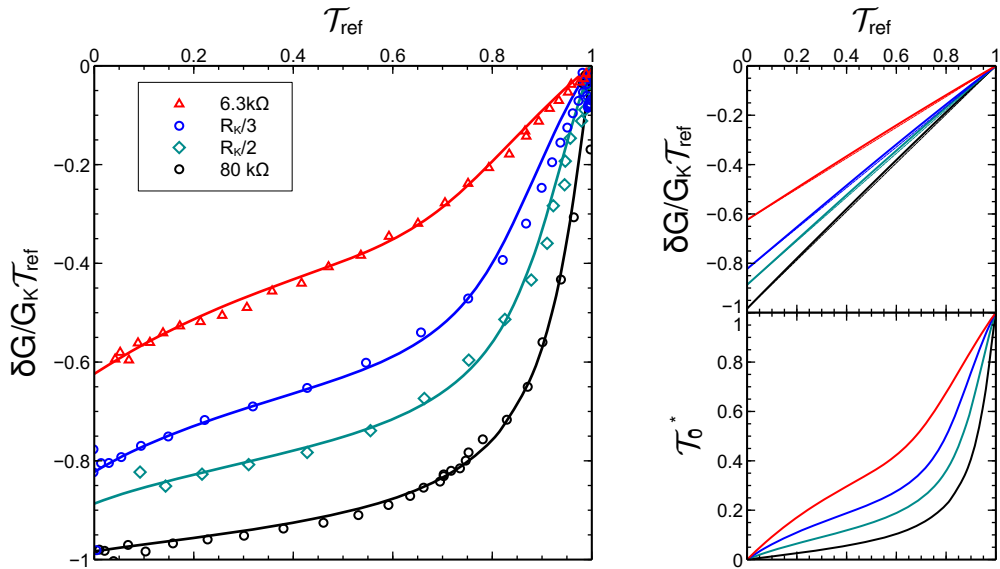

FIG. 3. Left: Relative DCB reduction of conductance of a QPC as a function of a reference transmission $\mathcal{T}_{\text{ref}}$ of the QPC obtained using the Landauer formula in a situation where DCB is suppressed. Symbols are data from Fig. 3 of Ref. [18] and lines are adjustments using our theory. The curves are obtained combining the two right graphs. Top right: Relative reduction of conductance due to generalized dynamical Coulomb blockade for energy-independent transmission given by Eqs. (24) and (27) and the parameters of Table I. Because of the self-consistency of the phase fluctuations these curves very slightly deviate from the straight line (shown as dashes). These straight lines would generalize the $(1 − \mathcal{T}_0)$ Fano factor reduction of the tunnel limit of DCB previously known to be valid only in the case of weak reduction of conductance. Bottom right: Modified transmission of the QPC due to a change of the quantum capacitance, calculated using the parameters of Table I.

**discussion**

We find that our theory can well adjust the data for the full range of transmission and environmental impedance explored in the experiments of Pierre and coworkers, including for the strongest DCB with $R \simeq 3.1 R_K$. These adjustements illustrate the crucial role (Fig. 3, bottom right panel) played by the DCB of the quantum capacitance introduced in Sec. II and improve significantly on the approximate phenomelogical adjustment previously proposed in Refs. [52, 18] (see Supplemental Material).

We observe that the Gaussian approximation for the correlators seems to be sufficient to explain the whole parameter space explored in the experiments of Pierre and coworkers. In such a circuit without any quasi-isolated electrode where charging effects could occur, the parameters that would be needed in order to observe a significant departure from our predictions (and hence requiring to consider higher order cumulants [51, 50]) is thus an open question.

The quality of the adjustments shown in Fig. 3 may even be surprising given the extremely sketchy model used for the QPC. We notably (i) use a 0-D modeling of the QPC that neglects the spacial distribution of charges inside the channel [59, 60], (ii) we assume a potential barrier of variable height but with a fixed shape, and (iii) that the geometrical capacitances are independent of the gate voltage which is known to be not strictly valid in a QPC around the pinch-off [62]. These latter two approximations are both expected to break down when the channel gets depleted. A more realistic model (e.g. [63]) taking these effects into account would furthermore eliminate the need for the high barrier energy cutoff that necessitated an extra adjustable parameter in the voltage dependence of the electrostatic balance Eq. (32), as discussed above. Once this done, one should be able to relate our theory to the predicted [12, 10, 11, 13] and observed [18] scaling law of the conductance with the bias voltage. Although we leave this for future works, we believe the non-universal energy scale contained in this scaling should somehow be related to the non-universality we find in the DCB-induced electrostatic modification of the transmission.

Finally, the analysis developped in this section is merely a special case of the fairly general one developed in Sec. II which should apply equally well to other systems than the QPC in an ohmic environment considered here. For instance it could address DCB caused by a high impedance resonant mode which can be implemented experimentally [21] and which has not been considered theoretically so far at finite transmission. As well, it could describe a quantum dot (treated as a single resonant level) by using the Breit-Wigner scattering matrix instead of Eq. (20) for the QPC considered here. As always, for our approach to be consistent, the dwell time in the structure shall be shorter than the typical timescale of electromagnetic fluctuations. Away from the level's resonance, the energy dependence of the Breit-Wigner scattering matrix vanishes so that this condition is always verified whatever the transmission of the dot. In that case the dot operates in the cotunneling regime and behaves as an effective single tunnel barrier. Close to the resonance however, this approach is restricted to relatively open dots. For more closed dots, instead of using the global scattering matrix of double-barrier systems, with some more work one should be able to consider the individual scattering matrices of the two barriers and the (short) propagation in-between in presence of the fluctuating fields. We believe the measurements performed in carbon nanotubes and reported in Refs. [16, 64] could be analyzed that way.

## IV. CONCLUSIONS, SCOPE AND GENERALIZATIONS

We have considered the effect of quantum voltage fluctuations in a short coherent conductor embedded in an arbitrary external circuit. We have shown how fluctuations in such circuit can be consistently incorporated at the minimal level into the Landauer-Büttiker formalism. This is achieved by assuming fluctuations are all Gaussian and taking into account the back-action of the conductor itself on the rest of the circuit, including in non-equilibrium situations. Our approach synthesizes the LB and $P(E)$ theories and formally solves the electrodynamics of quite general quantum electronic systems. This analysis shows how low frequency electromagnetic modes generically mediate interactions by correlating successive electron transfer and make the system non-local and non-linear. It also unveils a new effect whereby voltage fluctuations combined with the energy-dependence of the scattering matrix induce a change of in local electronic DOS in the scatterer which may have a large impact on its transport properties.

Detailed predictions can be derived from the general description provided one knows the form of the energy-dependent scattering matrix. We discuss the specific case of a single-mode quantum point contact of arbitrary transmission. In that case we explicitly work out how phase fluctuations reduce the DOS of the device and change its overall transmission in a predictable, albeit non-universal way. Once this effect taken into account we widely generalize the predictions of the usual dynamical Coulomb blockade theory (a.k.a. $P(E)$ theory), to arbitrary transmission and arbitrary environment impedance, in which the conductance is not a direct measurement of the (modified) transmission, unlike in the standard Landauer-Büttiker scattering theory. Using this approach we are able to account for experimental data of QPCs in series with resistances up to a few $R_K$ for which no fully predictive theory existed.

Overall, our derivation involves only concepts familiar to electronics engineers and it can be easily implemented numerically for any given model of scattering matrix and external impedance for which it should apply equally well.

Given the similarity of our approach with the usual *P(E)* theory, many results previously obtained for tunnel junctions in that framework should be amenable to a generalization in short coherent conductors. In particular, results for photon-assisted transport [65, 40, 21, 66], or results regarding the properties of the radiation emitted by tunnel junctions [67, 68, 69] could be extended to finite transmissions. Conversely, questions addressed in the LB formalism, like e.g. heat transport [70, 71], can now be revisited taking into account the effect of fluctuations.

Finally, by contributing to the better general understanding of quantum circuit this work should open up new perspectives in the engineering of their properties. One could for instance design an electromagnetic environment that implements some type of (passive or active) feedback on a coherent conductor [72], or structure it to achieve different properties at different frequencies.

### ACKNOWLEDGEMENTS

The authors acknowledge useful discussions and inputs from C. Glattli, H. le Sueur, C. Mora, P. Roche, I. Safi and the Quantronics Group. We are specially thankfull to A. Anthore, S. Jezouin, F. Parmentier and F. Pierre for in-depth discussions and their assitance for the comparison with their experimental results.

This work received funding from the European Research Council under the European Union's Programme for Research and Innovation (Horizon 2020) / ERC grant agreement number [639039], the ANR AnPhoTeQ research contract and the PALM-P2IO NDS-NbSi project.

# Supplemental Material for
# Interacting electrodynamics of short coherent conductors in quantum circuits

C. ALTIMIRAS    F. PORTIER    P. JOYEZ

(Dated: April 28, 2016)

**DERIVATION OF EQ. (12).**

When one knows all noise sources (assumed to be independent current sources) and all the impedances in a linear circuit, the voltage fluctuations in the circuit are entirely determined by the linear dynamics. Such results can be found in the literature [73, 29], but their status with respect to the thermal equilibrium assumptions is not always clear. We show here that it is a straight-forward consequence of linear electromagnetic response solely and that it is valid out-of-equilibrium.

In a linear circuit, the vector of voltages $[V(t)]$ at the ports of the circuit is related to the vector of currents $[I(t)]$ at the ports through the time-domain impedance matrix $\mathbf{Z}$:

$$[V(t)] = \int_0^{+\infty} \mathbf{Z}(\tau) \cdot [I(t-\tau)] d\tau = \int_{-\infty}^{+\infty} \mathbf{Z}(\tau) \cdot [I(t-\tau)] d\tau \qquad (Z(t<0)=0 \text{ by causality})$$

where $\mathbf{Z}(t) = \int d\omega \mathbf{Z}(\omega) e^{-i\omega t}$ and $\mathbf{Z}(\omega) = \mathbf{Y}^{-1}(\omega)$ is the inverse of the admittance matrix considered in the main text. Then, the voltage correlation matrix is given by

$$\begin{aligned}
\mathbf{S}_{VV}(t) &= \langle [V(t)] \cdot [V(0)]^T \rangle \\
&= \left\langle \left( \int d\tau \mathbf{Z}(\tau) \cdot [I(t-\tau)] \right) \cdot \left( \int d\tau' \mathbf{Z}(\tau') \cdot [I(0-\tau')] \right)^T \right\rangle \\
&= \int d\tau \int d\tau' \mathbf{Z}(\tau) \cdot \langle [I(t-\tau+\tau')] \cdot [I(0)]^T \rangle \cdot \mathbf{Z}^T(\tau') \\
&= \int d\tau \int d\tau' \mathbf{Z}(\tau) \cdot \mathbf{S}_{II}(t-\tau+\tau') \cdot \mathbf{Z}^T(\tau') \\
&= \int d\tau \int d\tau' \left( \int d\omega' \mathbf{Z}(\omega') e^{-i\omega'\tau} \right) \cdot \left( \int d\omega \mathbf{S}_{II}(\omega) e^{-i\omega(t-\tau+\tau')} \right) \cdot \left( \int d\omega'' \mathbf{Z}(\omega'') e^{-i\omega''\tau'} \right)^T \\
&= \int d\tau \int d\tau' \int d\omega' \int d\omega \int d\omega'' Z(\omega') e^{-i(\omega'-\omega)\tau} \cdot \mathbf{S}_{II}(\omega) e^{-i\omega t} \cdot (Z(\omega'') e^{-i(\omega''+\omega)\tau'})^T \\
&= \int d\omega' \int d\omega \int d\omega'' \mathbf{Z}(\omega') \delta(\omega-\omega') \cdot \mathbf{S}_{II}(\omega) e^{-i\omega t} \cdot \mathbf{Z}^T(\omega'') \delta(\omega+\omega'') \\
&= \int \mathbf{Z}(\omega) \cdot \mathbf{S}_{II}(\omega) \cdot (\mathbf{Z}(\omega))^\dagger e^{-i\omega t} d\omega
\end{aligned}$$

where T and † denote transposition and Hermitian conjugation (transpose-conjugate), repectively. Finally since the phase is a time-integral of the voltage and considering the parallel connection shown in Fig. 1c, we arrive at Eq. (12).

**ZERO-TEMPERATURE COMMON MODE CURRENT NOISE**

In the case of a QPC with the Kemble transmission probability (Eq. (21)) considered in sec. III., the zero-temperature and zero-voltage limit of $S_{I_3 I_3}$ (Eqs. (15), (5)) can be evaluated analytically. This gives

$$S^0_{I_3 I_3}(t,V) = S^{0R}_{I_3 I_3}(t,E_0) + \cos\frac{eVt}{\hbar} e^{J(t)} S^{0J}_{I_3 I_3}(t,E_0)$$

where $V$ is the bias voltage,

$$S^{0R}_{I_3 I_3}(t, V) = \frac{e^2}{\hbar^2}\Bigg(2\, i\, \pi\, \delta'(t) + 2\, e^{-\pi\omega t}\, \pi\left(-4\, \beta_{-1}\left(-i\,\omega\, t, \frac{1}{2}\right) + 4\, e^{2\pi\omega t}\, i\, \beta_{-1}\left(i\,\omega\, t + \frac{1}{2}, \frac{1}{2}\right) - \beta_{-\frac{1}{\lambda}}(-i\,\omega\, t,$$

$$0) + 2\, \beta_{-\frac{1}{\lambda}}\left(-i\,\omega\, t, \frac{1}{2}\right) + 2\, \beta_{-\lambda}\left(-i\,\omega\, t, \frac{1}{2}\right) + 2\, i\, \pi\, (\coth(\pi\,\omega\, t) + 1)\Bigg)\,\delta(t) - 2\Bigg(2\, e^{-\pi\omega t}\, \frac{\lambda^{-i\omega t}}{\omega t}\left(i\, \beta_{-\frac{1}{\lambda}}(-i\,\omega\, t, 0) + \pi\, (\coth(\pi\,\omega\, t) + 1)\right) + \frac{1}{(\omega t)^2} + i\,\pi\,\frac{\lambda^{-i\omega t}}{\omega t}\,\delta(t)\Bigg)\,_2F_1(1, -i\,\omega\, t; 1 - i\,\omega\, t; -\lambda) + 2\Bigg(\pi\frac{\lambda^{i\omega t}}{\omega t}\mathrm{csch}(\pi\,\omega\, t) +$$

$$2\,\pi^2\,(\coth^2(\pi\,\omega\, t) - \tanh^2(\pi\,\omega\, t)) + e^{-\pi\omega t}\,\pi\,\frac{\lambda^{-i\omega t}}{\omega t} + e^{-\pi\omega t}\, i\,\frac{\lambda^{-i\omega t}}{\omega t}\left(2\,\beta_{-1}\left(-i\,\omega\, t, \frac{1}{2}\right) + \beta_{-\frac{1}{\lambda}}(-i\,\omega\, t, 0) - 2\,\beta_{-\frac{1}{\lambda}}\Big(-i\,\omega\, t,$$

$$\frac{1}{2}\Big)\Bigg) + 2\,\beta_{-\frac{1}{\lambda}}\left(-i\,\omega\, t, \frac{1}{2}\right)\left(-\beta_{-\frac{1}{\lambda}}(i\,\omega\, t, 0) + \beta_{-\frac{1}{\lambda}}\Big(i\,\omega\, t, \frac{1}{2}\Big) + \beta_{-\frac{1}{\lambda}}(i\,\omega\, t + 1, 0)\right) + 2\,\beta_{-\lambda}\left(-i\,\omega\, t, \frac{1}{2}\right)\beta_{-\lambda}\Big(i\,\omega\, t,$$

$$\frac{1}{2}\Big) - 2\,\beta_{-1}\left(-i\,\omega\, t, \frac{1}{2}\right)\left(-\beta_{-\frac{1}{\lambda}}(i\,\omega\, t, 0) + \beta_{-\frac{1}{\lambda}}\Big(i\,\omega\, t, \frac{1}{2}\Big) + \beta_{-\frac{1}{\lambda}}(i\,\omega\, t + 1, 0) + \beta_{-\lambda}\Big(i\,\omega\, t, \frac{1}{2}\Big)\right) + 2\,e^{\pi\omega t}\,\beta_{-1}\Big(i\,\omega\, t + \frac{1}{2},$$

$$\frac{1}{2}\Big)\left(\frac{\lambda^{-i\omega t}}{\omega t} + e^{\pi\omega t}\, i\left(-\beta_{-\frac{1}{\lambda}}(i\,\omega\, t, 0) + \beta_{-\frac{1}{\lambda}}\Big(i\,\omega\, t, \frac{1}{2}\Big) + \beta_{-\frac{1}{\lambda}}(i\,\omega\, t + 1, 0) + \beta_{-\lambda}\Big(i\,\omega\, t, \frac{1}{2}\Big)\right)\right) - 2\,e^{-2\pi\omega t}\,\beta_{-\frac{1}{\lambda}}\Big(\frac{1}{2} - i\,\omega\, t,$$

$$0\Big)\,\beta_{-\lambda}\Big(\frac{1}{2} - i\,\omega\, t, 0\Big) + \frac{4\,i\,\pi\,\beta_{-\lambda}\Big(\frac{1}{2} - i\,\omega t, 0\Big)}{1 + e^{2\pi\omega t}} + 2\,i\,\pi\,\beta_{-\frac{1}{\lambda}}(-i\,\omega\, t, 0)\,(\coth(\pi\,\omega\, t) - 1) + e^{-\pi\omega t}\,\pi\,\frac{\lambda^{-i\omega t}}{\omega t}\coth(\pi\,\omega\, t) -$$

$$2\,i\,\pi\,\beta_{-\frac{1}{\lambda}}\Big(\frac{1}{2} - i\,\omega t, 0\Big)(\tanh(\pi\,\omega t) - 1) - \frac{1}{(\omega t)^2}\Bigg)\Bigg)$$

and

$$S^{0J}_{I_3 I_3}(t, V) = \frac{e^2}{\hbar^2}\Bigg(-\frac{2}{(\omega t)^2} - 2\,\pi\,\frac{\lambda^{-i\omega t}}{\omega t}(\lambda^{2i\omega t} + 1)\,\mathrm{csch}(\pi\,\omega\, t) - 16\,\pi^2\,\coth(2\,\pi\,\omega\, t)\,\mathrm{csch}(2\,\pi\,\omega\, t) + 2\,i\,\pi\,\delta'(t)\,e^{-i\omega t|\log(\lambda)|} +$$

$$\delta(t)\left(\frac{2\,i\,\pi}{\omega t} + 2\,e^{-\pi\omega t}\,\pi\,\beta_{-\frac{1}{\lambda}}(-i\,\omega\, t, 0) - 4\,i\,\pi^2\,\mathrm{csch}(\pi\,\omega\, t) + 2\,i\,\pi\,\frac{\lambda^{-i\omega t}}{\omega t}(_2F_1(1, -i\,\omega\, t; 1 - i\,\omega\, t; -\lambda) - 1)\right) +$$

$$\left(4\,e^{-\pi\omega t}\,\frac{\lambda^{-i\omega t}}{\omega t}\left(i\,\beta_{-\frac{1}{\lambda}}(-i\,\omega\, t, 0) + \pi\,\coth(\pi\,\omega\, t) + \pi\right) + \frac{2}{\omega t^2}\right)\,_2F_1(1, -i\,\omega\, t; 1 - i\,\omega\, t; -\lambda) - 2\,i\,\frac{\lambda^{-i\omega t}}{\omega t}\,\beta_{-\frac{1}{\lambda}}(-i\,\omega\, t,$$

$$0)(\coth(\pi\,\omega\, t) - 1)(2\,\pi\,\omega\, t\,\lambda^{i\omega t} + \sinh(\pi\,\omega\, t)) + \beta_{-\lambda}\Big(\frac{1}{2} - i\,\omega\, t, 0\Big)\left(4\,e^{-2\pi\omega t}\,\beta_{-\frac{1}{\lambda}}\Big(\frac{1}{2} - i\,\omega\, t, 0\Big) + 4\,i\,\pi\,(\tanh(\pi\,\omega\, t) -$$

$$1)\right) + 4\,i\,\pi\,\beta_{-\frac{1}{\lambda}}\Big(\frac{1}{2} - i\,\omega t, 0\Big)(\tanh(\pi\,\omega t) - 1)\Bigg)$$

with $\omega = \frac{\Omega}{2\pi}$, $\lambda = \frac{\mathcal{T}_0}{1-\mathcal{T}_0}$, $\beta$ the Euler beta function, and $_2F_1$ the Gauss hypergeometric function.

## COMPARISON WITH THE PHENOMENOLOGICAL SCALING LAW

In Refs. [52, 18] it was shown that the measured relative reduction of the conductance approximately followed a phenomenological scaling law without any adjustable parameter (see e.g. Eq. (1) in [18] written here using our notations):

$$\frac{\delta G}{G_K \mathcal{T}_{\mathrm{ref}}} = \frac{G_{\mathrm{phenom}}(\mathcal{T}_{\mathrm{ref}}) - G_K \mathcal{T}_{\mathrm{ref}}}{G_K \mathcal{T}_{\mathrm{ref}}} \sim \frac{1 + F(\omega = 0, J, V = 0, T)}{1 + \mathcal{T}_{\mathrm{ref}} F(\omega = 0, J, V = 0, T)} - 1 \qquad (36)$$

with the function $F$ given in Eq. (25) of the article. This scaling law was shown to be related to one obtained by the mapping of the system onto an impurity in a TLL [11], with the reference transmission defined at high energy (see supplementary note 3 in [18]). In Fig. S1 we compare the above phenomenological scaling with the experimental data and the predictions of our theory. As could be expected, we observe that although the parameter-free phenomenological scaling captures some general features of DCB in this system, it is less accurate than our theory (with adjustable parameters), missing notably the inflexion point at $\mathcal{T}_{\mathrm{ref}} \sim 0.25$ clearly visible in the datasets with the two lowest resistances.

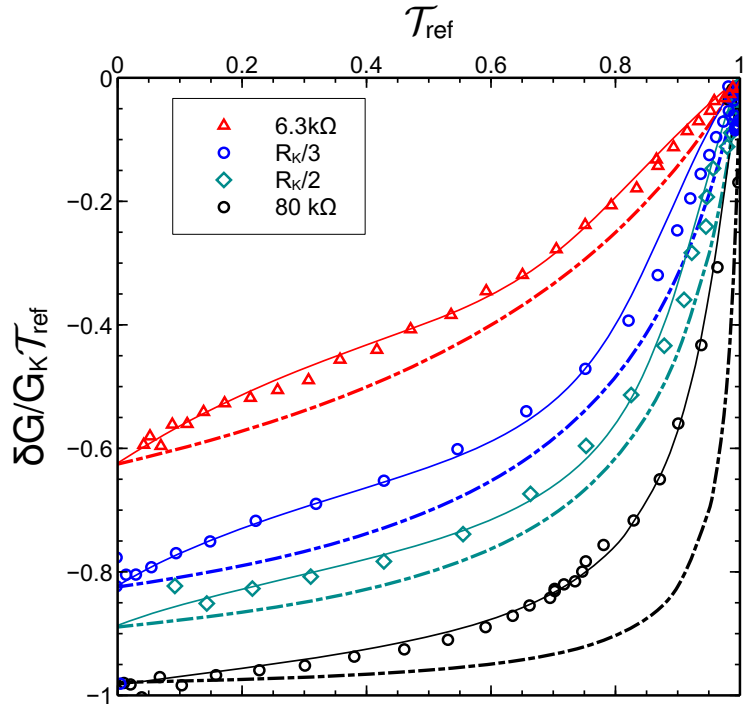

FIG.. Symbols are experimental values of the relative reduction of the conductance due to DCB as a function of the reference transmission of the QPC in absence of DCB. Thin solid lines are the adjustments shown in the main article, while thick dash-dotted lines are the predictions of the approximate phenomenological scaling law Eq. (36)